\documentclass[12pt,preprint]{aastex}

\input epsf.tex 
\def\kmsec{\mbox{km~s$^{\rm -1}$}}
\def\teff{\mbox{$T_{\rm eff}$}}
\def\logg{\mbox{log $g$}}
\def\vt{\mbox{$v_{\rm t}$}}
\def\BmV0{\mbox{$(B-V)^{\rm o}$}}
\def\VmK0{\mbox{$(V-K)^{\rm o}$}}
\def\MV0{\mbox{$M_{\rm V}^{\rm o}$}}

\def\Msun{\mbox{$\mathcal{M}_{\sun}$}}
\def\etal{\mbox{et al.}}
\def\eg{\mbox{e.g.}}

\def\deg{{$^{\circ}$}}

\begin{document}

\title{
Binary Blue Metal-Poor Stars: Evidence for AGB Mass Transfer
}

\author{
Christopher Sneden\altaffilmark{1,2},
George W. Preston\altaffilmark{2}, 
and
John J. Cowan\altaffilmark{3}
}

\altaffiltext{1}{Department of Astronomy and McDonald Observatory,
University of Texas, Austin, TX 78712; chris@verdi.as.utexas.edu}

\altaffiltext{2}{Carnegie Observatories, 813 Santa Barbara Street, 
Pasadena, CA 91101; gwp@ociw.edu}

\altaffiltext{3}{Department of Physics and Astronomy,
University of Oklahoma, Norman, OK 73019; cowan@mail.nhn.ou.edu}

\begin{abstract}

We present new abundance analyses of six blue metal-poor (BMP) stars
with very low iron abundances ([Fe/H]~$<$~--2), based on new 
high resolution echelle spectra.
Three are spectroscopic binaries and three have constant radial
velocities.
The chemical compositions of these two groups are very different, as
the binary BMP stars have large enhancements of carbon and 
neutron-capture elements that are products of $s$-process nucleosynthesis.
One star, CS~29497-030, has an extreme enhancement of lead, [Pb/Fe]~=~+3.7,
the largest abundance in any star yet discovered.
It probably also has an oxygen overabundance compared to the other BMP
stars of our sample.
The binary BMP stars must have attained their status via mass transfer
during the asymptotic giant branch (AGB) evolutions of their companion stars, 
which are now unseen and most likely are compact objects.
We have not found any examples of AGB mass transfer among BMP binaries with 
[Fe/H]~$>$~--2.

\end{abstract}


\section{INTRODUCTION}

Many Galactic globular clusters possess a small number of so-called
blue stragglers -- main sequence stars that are clearly bluer
and brighter than the turnoff stars.
Blue stragglers are generally believed to have formed from stellar 
mergers in the high density environments of these clusters 
(\eg, Stryker 1993\nocite{str93}; Sills \& Bailyn 1999\nocite{sil99}).

Stellar collisions are less frequent in the disk and halo by many orders 
of magnitude, so the substantial numbers of field blue stragglers identified 
and discussed by Preston, Beers, \& Shectman (1994)\nocite{pre94} and 
Preston \& Sneden (2000, hereafter PS00\nocite{pre00}) must be created 
by the only remaining mechanism, McCrea (1964)\nocite{mcc64} mass transfer.  
Of 62 BMP's investigated by PS00, 2/3 are in single-lined spectroscopic 
binaries with orbital periods in the approximate interval 
2~$<$~$P$(days)~$<$~4000.  
Analysis of the data indicates that half or more of the BMP stars are 
ordinary lower main sequence stars that gain mass from their companions 
during the post main-sequence evolution of the latter.  
The unseen companions are now some sort of compact white dwarf-like objects.  
We have argued that the remainder of the sample, including nearly all the 
single stars, are metal-poor stars of intermediate-age.  
A few of the single BMP stars may be the products of merger of the small 
fraction of binaries ($<$10\%) with initial periods less than 
$\sim$5~days (Vilhu 1982\nocite{vil82}).  
The intermediate-age population should also contribute its 
modest ($\sim$20 percent) share of spectroscopic binaries.   
Because we cannot imagine where in the Galaxy it would be possible
to preserve metal-poor gas for billions of years, we suggest that these 
intermediate-age stars may have been captured from metal-poor dwarf galaxies 
like the Carina dwarf spheroidal (Smecker-Hane \etal\ 1994\nocite{sme94}).

PS00 included determinations of Fe metallicities and abundance ratios 
of eight elements in most of the 62-star BMP sample, 
using co-added spectra of each star to increase the S/N.
Examples of individual and summed spectra are shown in Figure~1 of that paper.
Derived stellar metallicities were found to be uniformly spread 
over the range 0.0~$\gtrsim$~[Fe/H]~$\gtrsim$~--2.5. 
A normal halo abundance pattern was derived for stars with [Fe/H]~$<$~--0.5.
That is, the vast majority of BMP stars have enhancements of $\alpha$-elements:
[Mg/Fe]~$\approx$ [Ca/Fe]~$\approx$ [Ti/Fe]~$\approx$ +0.30~$\pm$~0.20.
For Fe-peak elements, [Sc/Fe]~$\approx$ [Cr/Fe]~$\approx$ 0.00~$\pm$~0.25.
And among neutron-capture ($n$-capture) elements, 
[Sr/Fe]~$\sim$ [Ba/Fe]~$\sim$~0.0 for stars with [Fe/H]~$>$ --2.0, but for
lower metallicities very large star-to-star scatter was found, especially
so for Ba.

PS00 made no attempt to compare abundance ratio systematics between 
the BMP binaries and those with no detectable radial velocity variations
(hereafter called RV-constant stars).
As part of an ongoing program to understand the differences in these two
BMP subclasses, we gathered new high resolution spectra of five binaries
and five RV-constant stars. 
Three stars in each of these groups have very low metallicities,
[Fe/H]~$\approx$ --2.1.
In this paper we concentrate on analyses of these six stars, concluding
that significant differences exist between the abundances of carbon and 
the $n$-capture elements of the binaries and the RV-constant stars. 
We argue that mass transfer from a former AGB companion star must have
created the binary BMP stars that we observe now.
We detect Pb in the spectrum of one of the BMP binaries, and find that
this star has the largest Pb abundance reported to date.
In \S 2 the observations are presented. 
In \S 3 the new radial velocities are given, along with revised orbits 
for the binary stars. 
We discuss the abundance analysis in \S 4, and the implications
of the abundances on the formation of binary BMP stars in \S 5.

\section{OBSERVATIONS AND REDUCTIONS}

Some basic data for the six program stars, taken from PS00, are given in 
Table~1.
The new observations were made with the du Pont echelle spectrograph
(Shectman 1984)\nocite{she84}, now outfitted with a Tektronics 
2048$\times$2048 CCD detector. 
A detailed description of the instrument and detector is
provided at the Carnegie Observatories website\footnote{
http://www.ociw.edu/lco/instruments/manuals/echelle/echelle.html}.
The spectral resolving power was the same as employed by PS00:
$R$~$\equiv$ $\lambda/\Delta\lambda$~$\approx$ 30,000.
Each star was observed between 2-4 separate times, with sub-integrations
combined to produce the final individual spectra.

All reductions leading to extracted, multiple-order spectra 
were made using standard IRAF\footnote{
IRAF is distributed by the National Optical Astronomy Observatories, which
are operated by the Association of Universities for Research in Astronomy,
Inc., under cooperative agreement with the National Science Foundation.} 
packages.  
The multiple exposures for each stellar spectrum were bias subtracted 
and median filtered to eliminate cosmic rays.
Generally small ($<$10\%) sky fluxes were removed by scaling and then
subtracting one of several sky observations made during each night, 
and the observations were divided by flat field images.  
One-dimensional extractions were obtained with the IRAF $apall$ task, 
subtracting at this time scattered light by measurement of inter-order flux.  
Hollow cathode Th-A comparison spectrum images, obtained before and after 
each stellar observation, were added and used to provide wavelength 
calibrations for the stellar observations.

\section{RADIAL VELOCITIES}

\begin{figure}
\epsscale{1.0}
\plotone{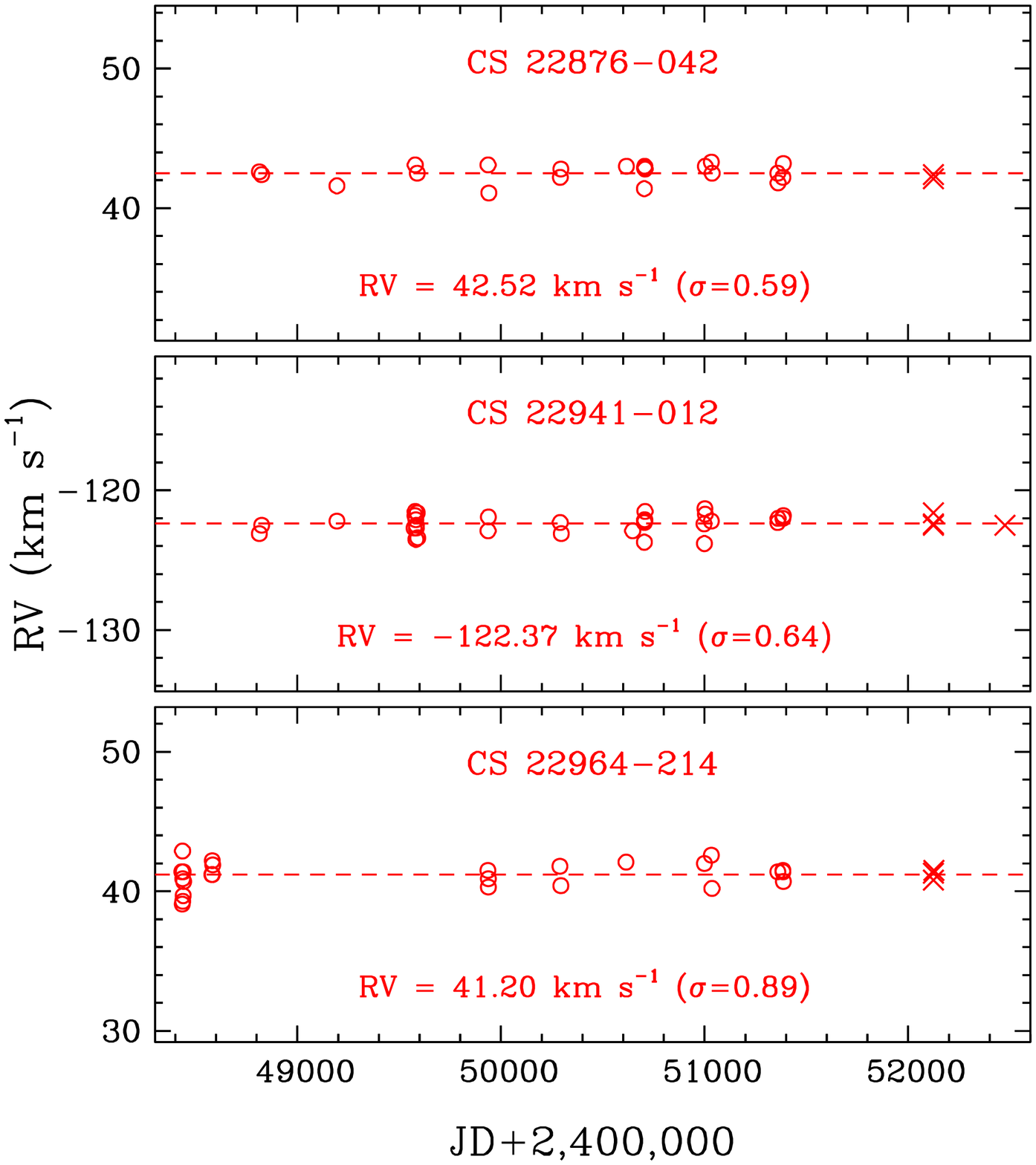}
\caption{
Velocities of the RV-constant stars as a function of Julian Date.
The velocities from PS00 are shown as open circles, and the
new velocities are shown as crosses.
The vertical range in each panel of this and the next figure
is 24~\kmsec.
\label{f1}}
\end{figure}

In PS00, we used a cross-correlation routine specifically designed
to determine BMP radial velocities from the 2D-FRUTTI spectra.
It was impractical to modify this procedure for use with the new CCD spectra.  
Instead, we used the IRAF $splot$ routine to measure wavelengths 
(hence velocities) for approximately 100 lines in the wavelength range 
3890-4590~\AA, selected from the line list used to derive the radial 
velocity of our BMP cross-correlation template star, CS~22874-009.  
Final radial velocities were straight means of these measures to
which heliocentric corrections were then applied.
The individual heliocentric radial velocities are given in Table~2.
Sample standard deviations calculated from measurements of 70 to 110 
lines in individual spectra are typically 0.3~\kmsec.  
These $\sigma$ values neglect systematic effects that vary from exposure 
to exposure, such as non-uniform slit-illumination and spectrograph flexure.  
Experience has shown (see PS00, Table~4) that velocity errors derived from 
multiple observations of RV-constant stars are two to four times larger than
the standard deviations for individual spectra.  
In Figure~\ref{f1} we show individual and mean radial velocities
of the RV-constant stars.
The new velocities are totally consistent with those of PS00.
In particular, the new RV means computed with both old and new RV measures 
are identical to ones just with the old RV's, and the $\sigma$ values are 
slightly decreased from those computed with just the older data.

\begin{figure}
\epsscale{1.0}
\plotone{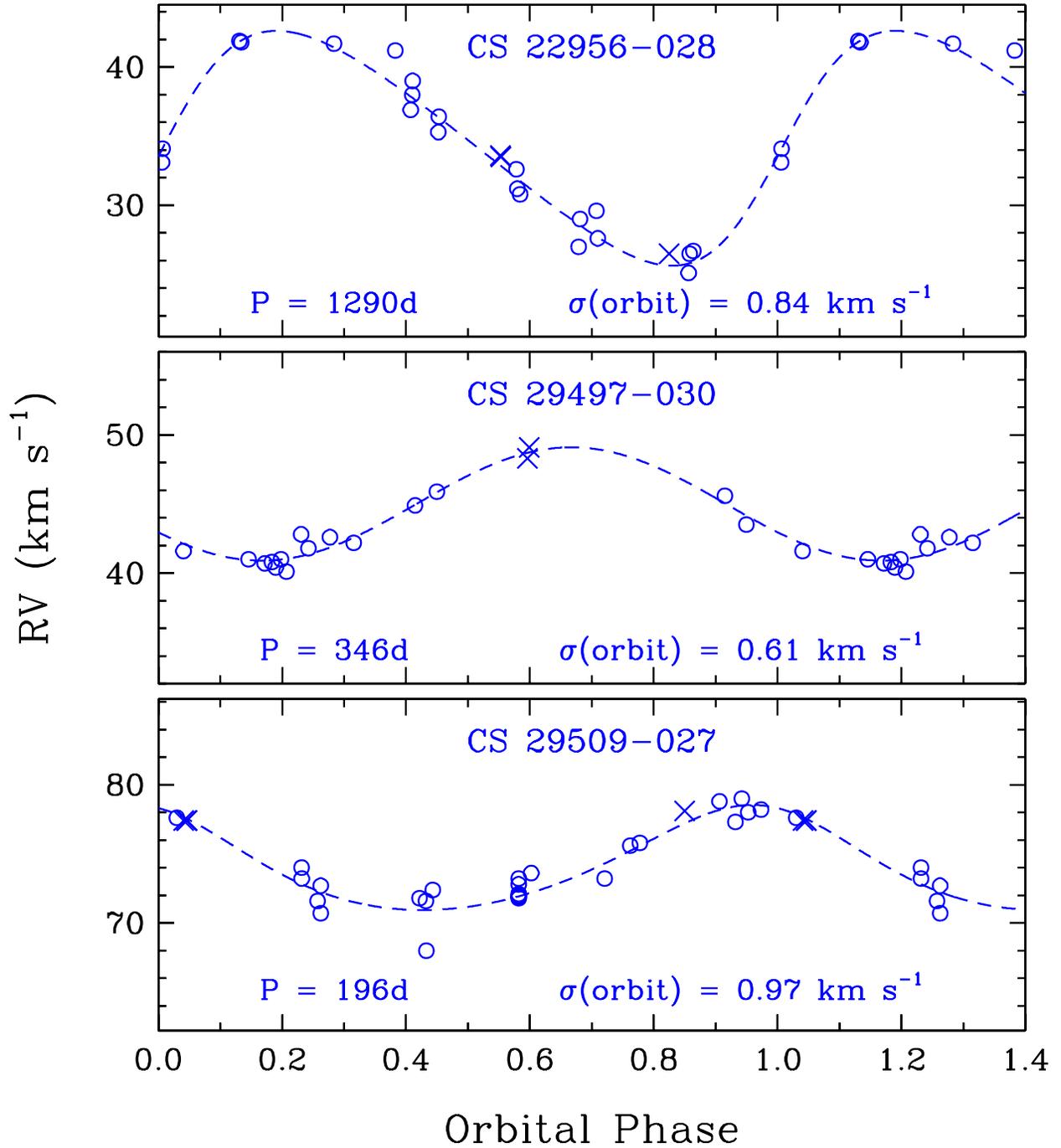}
\caption{
Velocities of the binary stars as a function of orbital phase.
The symbols and vertical ranges of the panels are as in the previous
figure.
The orbital solutions, plotted as solid curves, were obtained from
the parameters listed in Table~3.
\label{f2}}
\end{figure}

The new velocities for the binary stars have been combined with those
of PS00 to obtain improved orbital solutions.
The derived spectroscopic orbital parameters of the Julian Date of
periastron passage (JD$_0$), systemic velocity ($V_0$), velocity 
amplitude ($K_1$), eccentricity ($e$), longitude of periastron ($\omega$),
and period ($P$) are given in Table~3, along with the sample standard deviation
of the orbital solution ($\sigma_{\rm orb}$), and the number of velocities
used in the solution ($n$).
For CS~22956-028 no changes in the previous parameters were needed,
but new solutions were found for CS~29497-030 and CS~29509-027.
Most of the orbital parameter changes for these two stars are minor,
but the newly derived eccentricities are much smaller, moving them
into the the long-period but nearly-circular orbit category that
is over-represented among BMP binaries (see Figure~19 of PS00).
In Figure~\ref{f2} the individual velocities and the derived 
velocity curves are plotted as functions of orbital phase.

\section{ABUNDANCE ANALYSIS}

As was done by PS00, we re-binned the individual spectra for each star, 
and co-added them after shifting to a common (rest) wavelength scale.
This procedure yielded S/N ratios near echelle order centers for the final 
spectra of $\sim$80-100 at 6500~\AA, $\sim$60-90 at 5000~\AA, 
and $\sim$30-50 at 4000~\AA.
The spectra of PS00 were obtained with the 2D-FRUTTI detector 
that has very poor response in the yellow-red spectral region.
Thus the S/N values near 4000~\AA\ are comparable in old and new
spectra, but the S/N of the older data slowly decline with increasing
wavelength to useless levels beyond the Na~D lines near 5900~\AA.

Inspection of the co-added new CS~29497-030 spectrum revealed the presence
of CH G-band features from 4280-4325~\AA.
The combination of \teff~=7050~K and [Fe/H]~=--2.16 (PS00) should 
render G-band lines invisibly weak if [C/Fe]~$\sim$~0, so a substantial
overabundance of C was suggested from this detection.
Additional investigation led to detection of the G-band in all three of
the BMP binaries but in none of the RV-constant stars.
Noting also that abundances of Sr and Ba derived by PS00 were much higher
in the binaries than the RV-constant stars, we undertook a new abundance 
analysis of C, O, and $n$-capture elements in all six BMP stars.
Parameters of the atomic transitions employed in this study 
are given in Table~4.

The new and original spectra have comparable S/N in the blue, where the
majority of lines useful for model atmosphere determination are located.
Therefore, we did not repeat our previous full analysis of these stars.
Table~1 lists the atmospheric parameters \teff, \logg, \vt, and [Fe/H]
adopted from PS00.
To derive abundances in the present study we used model atmospheres 
from the Kurucz (1995)\nocite{kur95}\footnote{
see http://cfaku5.cfa.harvard.edu/} 
grid, interpolated between grid models by software kindly provided by
A. McWilliam (private communication).
Abundance calculations employed the current version of the LTE line
analysis code MOOG (Sneden 1973\nocite{sne73}).

\subsection{Carbon and Oxygen}

\begin{figure}
\epsscale{0.9}
\plotone{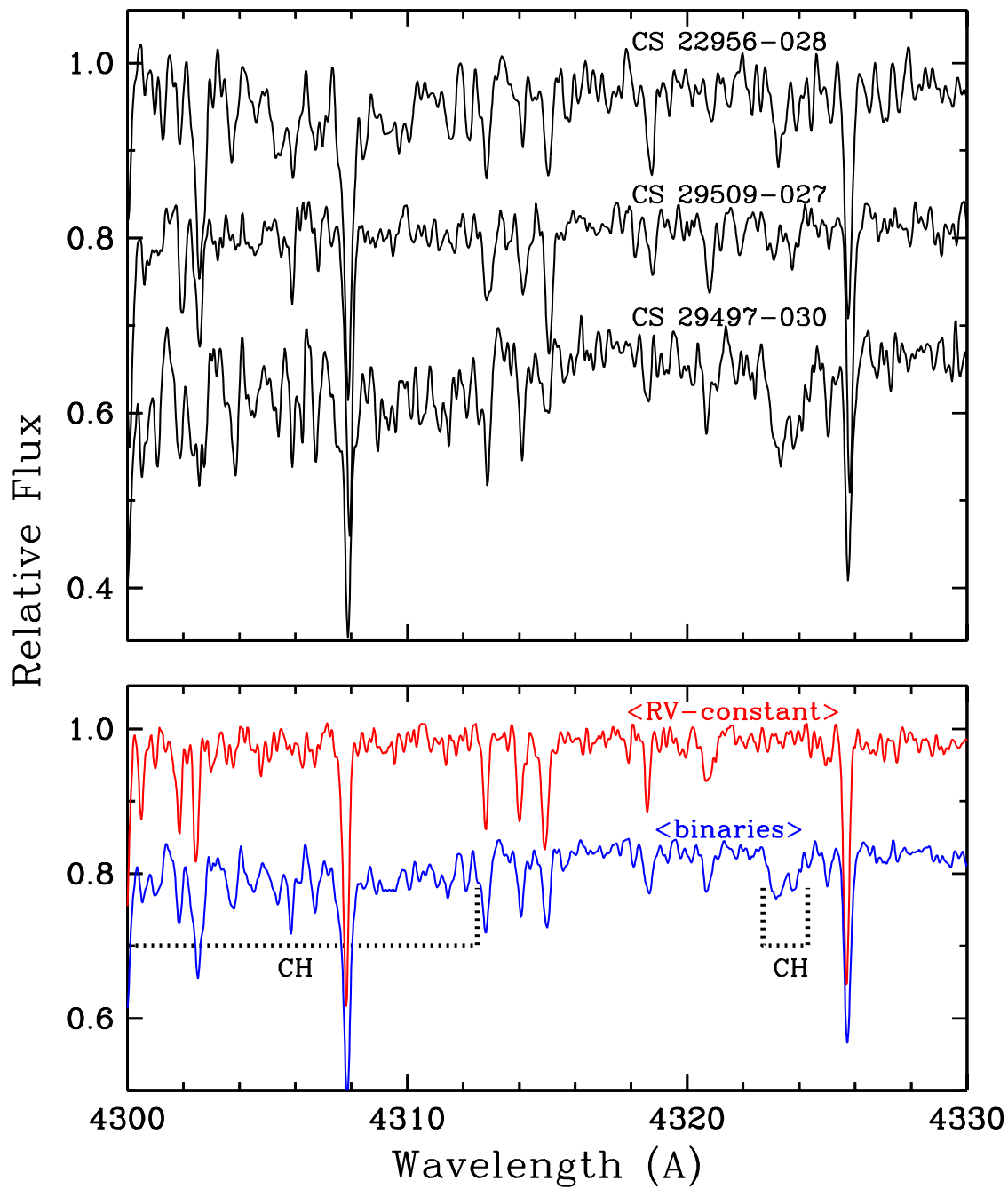}
\caption{
The CH G-band in the BMP program stars.
In the upper panel, individual spectra of the three binaries are shown
and in the lower panel the mean co-added spectra of the RV-constant
stars (in this and other figures labeled $<$RV-constant$>$)
and the binaries (labeled $<$binaries$>$) are shown.
In each panel the relative flux scale of the top spectrum is correct
and the other spectra are shifted downward by arbitrary additive
amounts for display purposes.
Regions of CH absorption are marked in the lower panel with dotted lines.
\label{f3}}
\end{figure}

In the upper panel of Figure~\ref{f3} we show the spectral region 
4300~$\leq$~$\lambda$(\AA)~$\leq$~4330 in the three BMP binaries.  
CH absorption, although generally weak, can be easily detected in all 
three stars. 
The CH lines reach more than 10\% depth in CS~29497-030.  
No CH absorption could be found in any of the RV-constant stars.
To demonstrate most clearly the CH contrast between the two groups of
BMP stars, we show in the lower panel of Figure~\ref{f3} the
averaged spectra of the RV-constant stars and the binaries.
These averaged spectra were formed by simple co-additions, giving each
star equal weight, after first re-binning the individual spectra and 
increasing the number of spectral points per spectral resolution 
element.

We determined C abundances for the stars from synthetic spectrum 
fits to their observed spectra in the wavelength range 
4280~$\leq$~$\lambda$(\AA)~$\leq$~4340.
The CH and atomic lines employed in these computations were culled
from the Kurucz (1995)\nocite{kur95} lists, as described in
Rossi \etal\ (1999\nocite{ros99}, and in preparation).
The solar photospheric abundance, employing this synthesis list,
the Holweger \& M{\"u}ller (1974)\nocite{hol74} solar model atmosphere, 
and the solar flux atlas of Kurucz \etal\ (1984)\nocite{kur84}, 
was log~$\epsilon_\sun$(C)~= 8.70~$\pm$~0.05.
The BMP relative abundances of C from the CH features, taken with respect
to this solar abundance, are given in Table~5.
They confirm that C is extremely enhanced in all three BMP binaries.
Unfortunately the upper limits on the RV-constant stars are uninformative,
since they would allow undetected large C abundances in those stars
as well.  

Since the BMP program stars all have similar atmospheric 
parameters (Table~1), we also derived mean CH-based C abundances 
for the two groups from their averaged spectra, using the 
following model stellar atmospheres.
For the RV-constant stars we adopted an average model with parameters 
(\teff, \logg, \vt, [Fe/H]) = (6900~K, 4.30, 2.3~\kmsec, --2.13), and for 
the binaries an average model with parameters 
(7000~K, 4.10, 1.9~\kmsec, --2.09).
The mean C abundance ([C/Fe]~= +1.76, Table~1) for the binaries determined 
from the co-added CH spectrum and the adopted mean model atmosphere 
is in excellent agreement with the mean of the individual abundances 
(fortuitously also +1.76), suggesting that this approach is valid.
The mean abundance for the RV-constant stars ([C/Fe]~$<$ +0.5)
is much lower than that suggested from the individual spectra
($<$+1.0); the increased S/N of the co-added spectrum allows a more
stringent upper limit to be determined.
This mean upper limit shows that there is no reason to suppose that
C enhancements exist in the RV-constant stars.

\begin{figure}
\epsscale{0.9}
\plotone{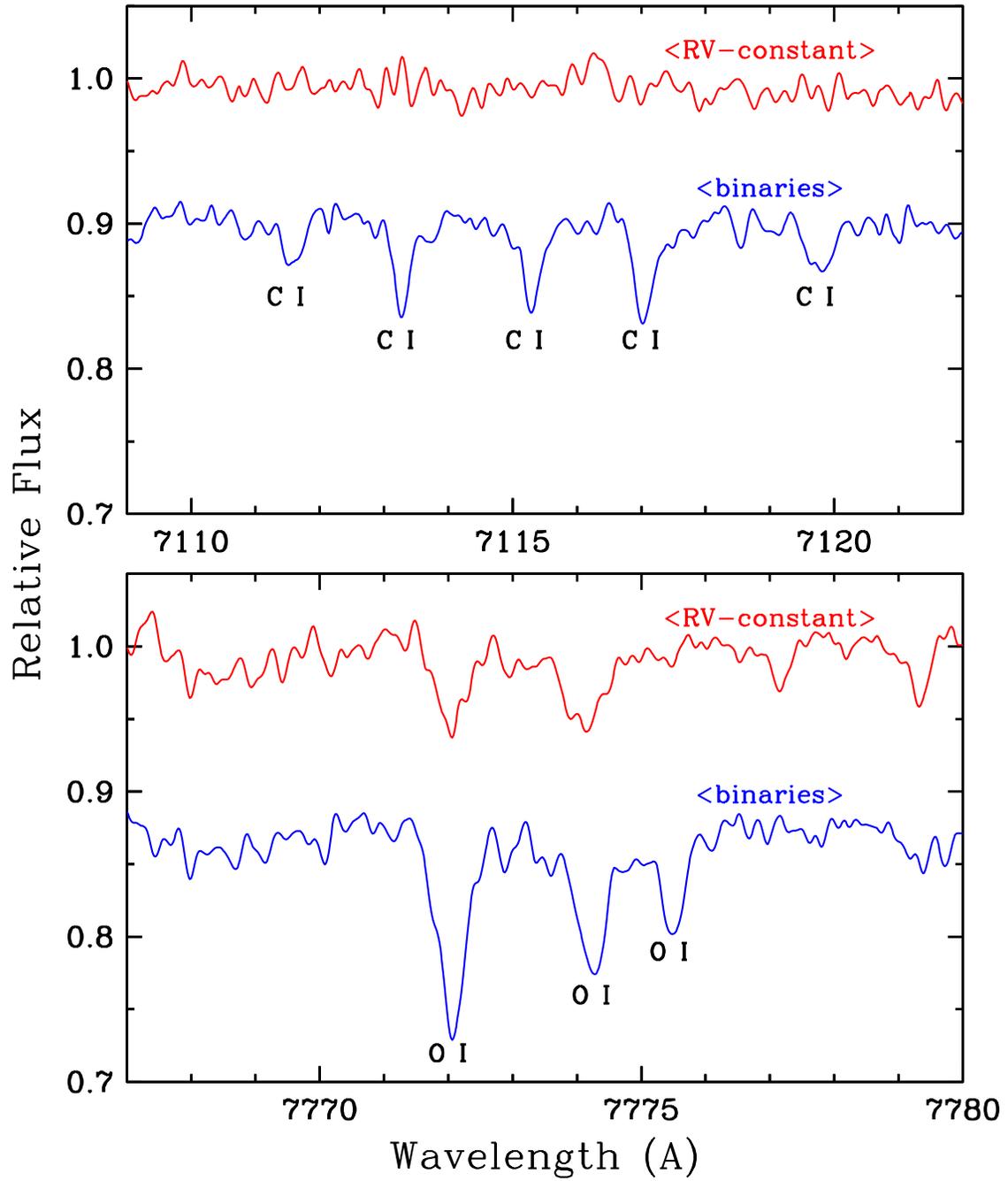}
\caption{
Spectra of \ion{C}{1} and \ion{O}{1} lines in the RV-constant and
the binary BMP stars.
\label{f4}}
\end{figure}

The reality of large C enhancements in BMP binaries is strengthened by
the detection of \ion{C}{1} lines in all three of these stars.
We show this in the upper panel of Figure~\ref{f4} with a display
of the mean spectra of RV-constant stars and binaries.
As many as seven \ion{C}{1} lines were detected on our spectra of the
binaries; none of these is apparent in the RV-constant stars.
Abundances and upper limits derived from these lines are listed in Table~5,
again taking the differences with respect to a solar abundance
of log~$\epsilon_\sun$(C)~= 8.42 derived from these same lines.
Both CH and \ion{C}{1} indicate very large overabundances, and the
difference between the abundances from the two sets of features for 
the same BMP binary exceeds 0.2~dex only in the case of CS~22956-028,
for which the \ion{C}{1} lines are barely detectable.

The O contents of two of the BMP binaries are similar to those of
the RV-constant stars.
In the lower panel of Figure~\ref{f4} we show the mean spectra
of the two groups, indicating that the \ion{O}{1} 7770~\AA\ triplet lines 
are $\sim$2-3 times stronger in the binaries, but the derived O abundances 
(Table~5) are the same within the errors for five of the six stars.  
The exception is CS~29497-030, whose O abundance is $\sim$3-4 times
larger than the other stars.  
This star's spectrum creates the difference in the mean \ion{O}{1} spectra 
and O abundances between the two groups.
The offset of CS~29497-030 appears to be real, but unfortunately 
no other O features are available on our spectra for confirmation.

\subsection{Strontium and Barium}

In PS00, Sr abundances were determined from the \ion{Sr}{2} 
4077, 4215~\AA\ resonance lines, and Ba abundances from just the \ion{Ba}{2}
4554~\AA\ resonance line.
We have repeated this analysis with the new spectra, but with the
increased S/N in the yellow-red we were able to detect other strong
\ion{Ba}{2} lines at 5853, 6141, and 6496~\AA.
Both \ion{Sr}{2} lines and the 4554~\AA\ \ion{Ba}{2} line were detected
in all six stars.
The abundances of Sr and Ba are listed in Table~5.
The average abundances derived from the mean spectra are consistent
with the means of abundances derived for each star.
The new Sr and Ba values are generally consistent with those determined by PS00.
The only exception is in Sr abundances of the BMP binaries, for which
the new abundances are $\approx$0.2~dex larger in the present analysis.
We attribute the difference to our use of synthetic spectrum computations
in the present analysis instead of the single-line EW analysis of PS00.

\begin{figure}
\epsscale{0.9}
\plotone{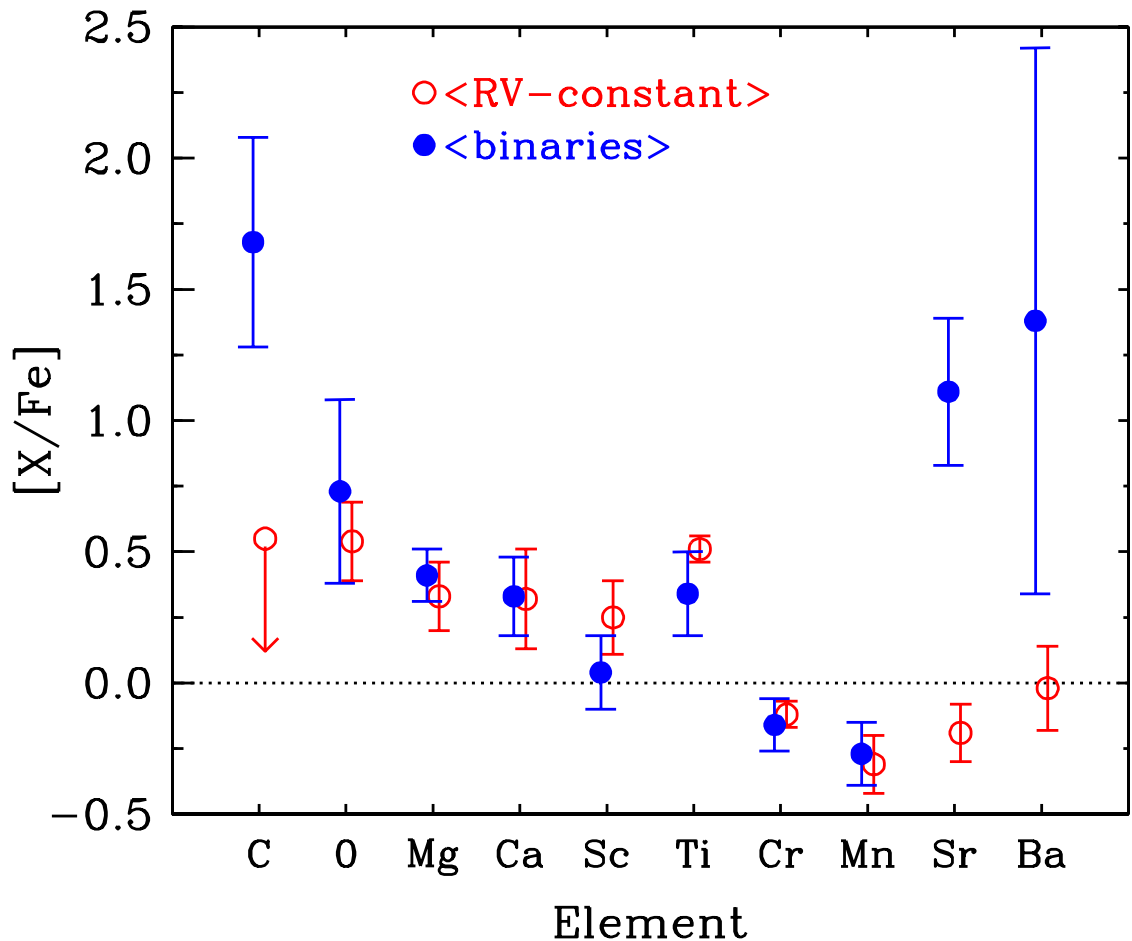}
\caption{
Average abundances of the BMP RV-constant and binary groups.
Abundances of C, O, Sr, and Ba are from this paper, and those of
the other elements are from PS00.
Small horizontal shifts have been introduced to the points to the left 
for the binaries, and to the right for the RV-constant stars, for display 
purposes.
Each abundance and its vertical range is the mean and the sample standard
deviation $\sigma$ of the values for each star in the group.
The true observational/analytical uncertainties are typically about
$\pm$0.15.
Therefore the very large ranges for C, Sr, and Ba for the binaries represent 
true star-to-star scatter, far beyond abundance measurement errors.
\label{f5}}
\end{figure}

It is clear that Sr and Ba generally have $>$1~dex abundance enhancements 
in the BMP binaries.
Since no other $n$-capture element transitions are detectable in all 
of the binary stars, we now illustrate the general abundance contrasts
between RV-constant stars and the binaries in Figure~\ref{f5}.
Newly-derived abundances and those from PS00 are combined in this
figure.
Note the very good agreement for the majority of the elements in the
two stellar groups.
Significant differences between them occur only for C, Sr, and Ba.

\subsection{Abundances in CS~29497-030}

\begin{figure}
\epsscale{0.9}
\plotone{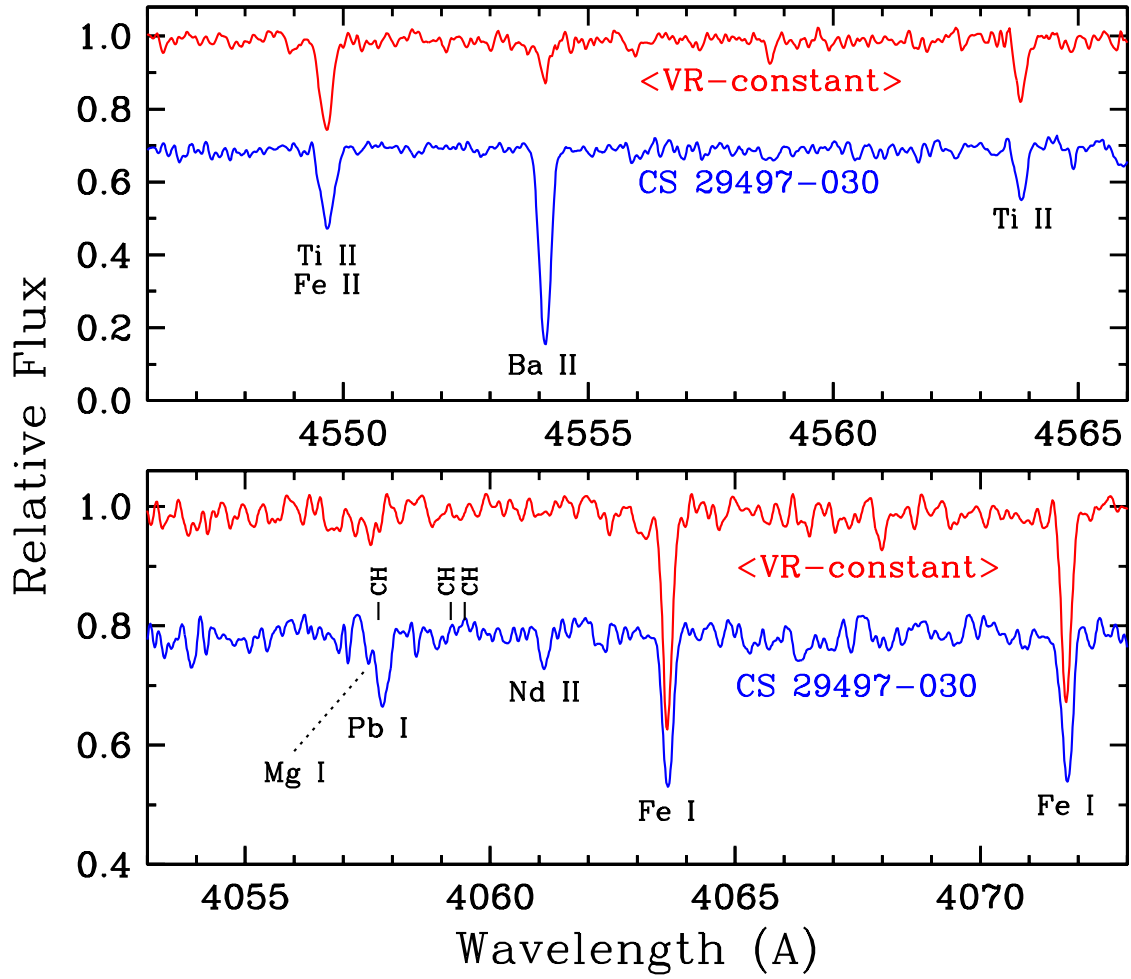}
\caption{
Comparison of the strongest \ion{Ba}{2} line (upper panel) and the only 
accessible \ion{Pb}{1} line (lower panel) in the mean RV-constant spectrum
and in CS~294097-030.
\label{f6}}
\end{figure}

The C and Ba abundances of the CS~29497-030 photosphere reach or exceed 
their values in the Sun, in spite of this star's very low metallicity.
Further examination of our new spectrum of this star revealed the 
presence of a few other $n$-capture transitions, most notably the 
4057~\AA\ line of \ion{Pb}{1}.
In the upper panel of Figure~\ref{f6} we show the contrast in 
\ion{Ba}{2} 4554~\AA\ line strengths between CS~29497-030 and the mean
RV-constant star spectrum, and in the lower panel the Pb feature.
Although this is the sole \ion{Pb}{1} line available on our spectrum, the
reality of the Pb absorption is not in doubt, because it can be also
detected on PS00's noisier spectrum of this star.

Abundances of La, Eu, and Pb were derived from synthetic spectrum
matches to the few available lines of these elements, and from the
EW of the single detected \ion{Nd}{2} line.
All the CS~29497-030 abundances from this work and from PS00 are
listed in Table~6.
The Nd and Eu abundances should be viewed with caution. 
The \ion{Eu}{2} 4129, 4205~\AA\ lines are barely detectable on our spectra.
However, the Eu abundance enhancement certainly is much less than that 
of Ba, and probably less than those of La and Nd.
The \ion{Nd}{2} 4061~\AA\ line should be the strongest Nd feature on
our spectra, but there are other potentially useful \ion{Nd}{2} lines
(\eg, 4109.5, 4012.2~\AA) that are undetectable on our spectra.
Absorption is seen at \ion{Nd}{2} 4303.6~\AA, but the surrounding CH 
contamination prevents derivation of an Nd abundance from this feature.
A spectrum of CS~29497-030 with better S/N would undoubtedly reveal 
the presence of many of these other lines.

The Pb abundance was derived taking into account the presence of
a \ion{Mg}{1} line that produces a small amount of absorption at 4057.52~\AA\
and slightly contaminates the \ion{Pb}{1} 4057.81~\AA\ line.
We searched for other relatively strong $n$-capture-element transitions,
but they could not be detected.
We also included some CH lines in our syntheses, and positions of three of the 
most prominent CH features are marked in the lower panel of Figure~\ref{f6}.
The CH lines at 4059.2 and 4059.5~\AA\ should be stronger than the
potential CH contaminant to the \ion{Pb}{1} line, but neither are
detectable in our spectrum of CS~29497-030.
Even with extreme relative C enhancements, the BMP binaries are too warm
to produce much CH absorption in this wavelength region.

\section{DISCUSSION}

The very large C and $n$-capture abundances in our three very low 
metallicity BMP binaries, combined with [Ba/Eu]~$>$~0 in at least 
one star, argue strongly for the creation of these anomalies by slow 
$n$-capture nucleosynthesis (the $s$-process) acting in He-burning 
(C-producing) zones of AGB stars.
These enhanced abundances cannot have been synthesized {\it in situ} due 
to the main-sequence (i.e., pre-AGB) evolutionary state of the BMP stars. 
Instead, the $s$-process abundance enhancements must be the result of 
mass transfer from the (previous AGB) companion stars.
In this section we will explore further the $s$-process in low metallicity
stars, focusing on CS~29497-030 and other Pb-rich stars, and also comment 
on the division between those BMP stars that have $s$-process 
overabundances and those that do not.

\subsection{BMP Stars and $s$-Process Nucleosynthesis}

The $s$-process is responsible for roughly one-half of all isotopes heavier 
than iron in the solar system.
It has been identified as the source of $n$-capture elemental
overabundances in low- and intermediate-mass (0.8--8 \Msun) AGB
stars (Busso, Gallino \& Wasserburg 1999).\nocite{bus99}
Previous studies have also demonstrated that the dominant source of 
neutrons for the $s$-process in these AGB stars is the 
$^{13}$C($\alpha$,n)$^{16}$O reaction, although some neutrons may also be 
produced in $^{22}$Ne($\alpha$,n)$^{25}$Mg reactions.
As a result of convective mixing of hydrogen into the helium burning region,
various concentrations of $^{13}$C may be produced in these stars.
Depending upon that amount of $^{13}$C, relatively large neutron fluxes
with respect to the iron ``seed'' abundance may occur.
With more and more neutrons increased enrichments of the heavier elements 
will occur.

For low metallicity stars ({\it i.e.}, those with [Fe/H]~$\lesssim$~--1)
the $s$-process can lead to large overabundances of lead with respect to
other $s$-process elements, such as Ba.
This point is not new to our work, as other relatively Pb-rich metal-poor 
stars have been discovered by several groups.
Most of these stars are so-called CH giants (\eg, Van Eck \etal\
2001, 2003\nocite{van01,van03}; Johnson \& Bolte 2002\nocite{joh02}), 
but an increasing number exist near the main-sequence turnoff region 
(\eg, Aoki \etal\ 2001, 2002b\nocite{aok01,aok02b}; Carretta \etal\ 
2002\nocite{car02}; Lucatello \etal\ 2003\nocite{luc03}).
To date, CS~29497-030 appears to be the reigning Pb abundance champion 
among the $s$-process-enriched metal-poor stars, both relatively 
([Pb/Fe]~= +3.7) and absolutely ([Pb/H]~=~+0.5).  
But it could be dethroned at any time through discovery of an even more
extreme example. 
The Pb enhancement of CS~29497-030 also should decline when the star's 
convective envelope expands as it ascends the RGB to become a CH giant.

\begin{figure}
\epsscale{0.9}
\plotone{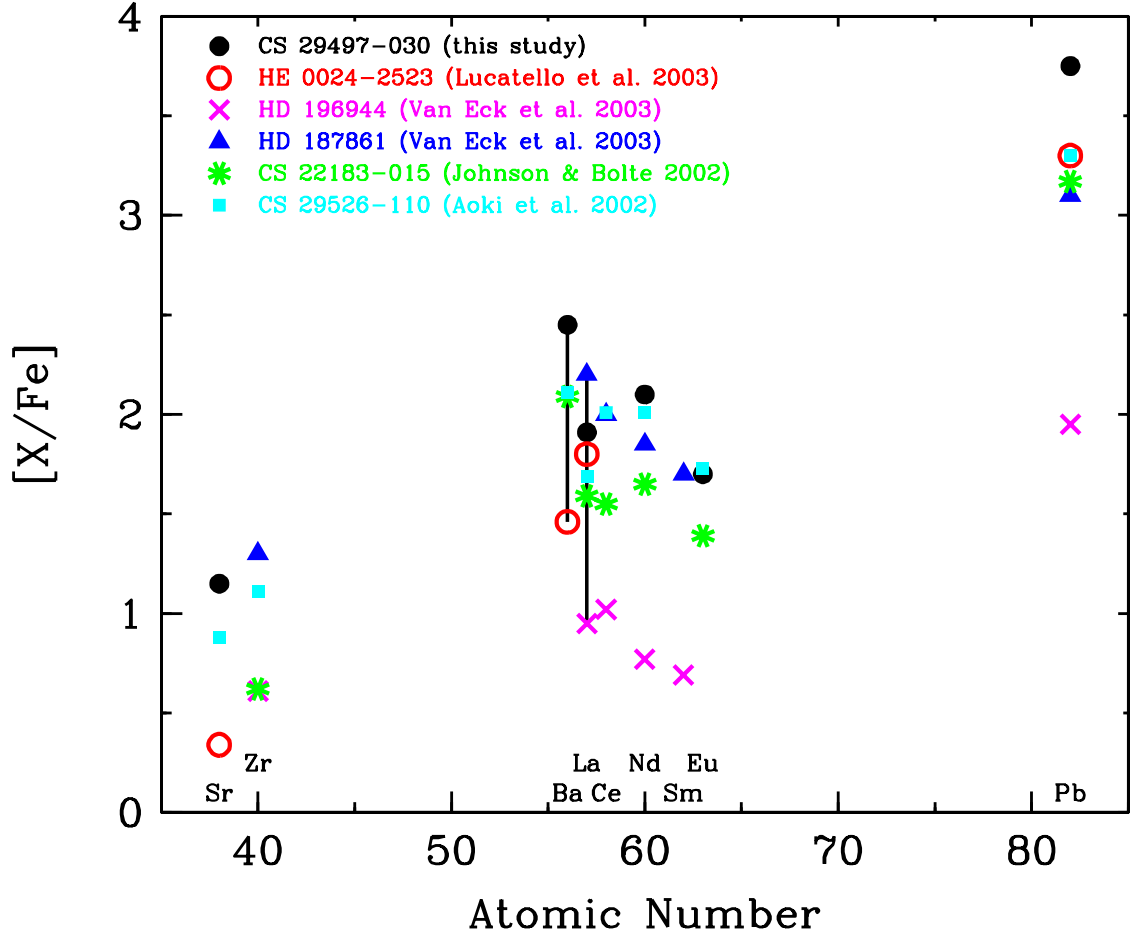}
\caption{
$n$-capture abundances in six extremely Pb-rich stars 
([Pb/(La or Ba)]~$\gtrsim$ +1.0).
The sources for the abundances are indicated in the figure legend.
The vertical lines connecting the Ba and La abundances signify that
mean abundances to be shown in the next figure have been computed after
renormalization of the abundance sets from other studies to the observed
Ba or La (or both) abundances of CS~29497-030.
\label{f7}}
\end{figure}

A more pertinent comparison with other Pb-rich stars is in the ratio
of the Pb abundance to those of lighter $n$-capture elements.
In Figure~\ref{f7} we display the $n$-capture abundances for
six stars with the largest Pb overabundances relative to Ba or La,
i.e, [Pb/(Ba or La)]~$\gtrsim$ +1.0.
Only elements observed in at least two stars are shown in this figure.
The [$<n$-capture$>$/Fe] values for these stars vary by more than one dex.
Additionally, different elements have been observed in different stars
(in the CH giants, some elements have transitions either too strong or too
blended for analysis; in the turnoff stars the lines of many element are
undetectably weak).
Therefore to more easily compare results from all the studies, we have 
renormalized the abundances of others to agree on average with our 
Ba and/or La abundances in CS~29497-030.
Taking all differences in the sense 
$\Delta$[X/Fe]~= [X/Fe]$_{\rm CS29497030}$ -- [X/Fe]$_{\rm other}$, 
the following offsets can be identified for the other data sets:
for HE~0024-2523 (Lucatello \etal\ 2003\nocite{luc03}), 
$\Delta$[Ba/Fe]~= +0.99 and $\Delta$[La/Fe]~= +0.11 for a mean of +0.55;
for HD~196944 (Van Eck \etal\ 2003\nocite{van03}), $\Delta$[La/Fe]~= +0.96;
for HD 187861 (Van Eck \etal), $\Delta$[La/Fe]~= --0.29;
for CS~22183-015 (Johnson \& Bolte 2002\nocite{joh02}),
$\Delta$[Ba/Fe]~= +0.36 and $\Delta$[La/Fe]~= +0.32 for a mean of +0.34;
and for CS~29526-110 (Aoki \etal\ 2002)\nocite{aok02b},
$\Delta$[Ba/Fe]~= +0.34 and $\Delta$[La/Fe]~= +0.22 for a mean of +0.28.
After application of these offsets, the abundances for each element from
the different studies were averaged.
The star-to-star scatter of the individual normalized values for an element
(other than Ba and La) was $\sigma$~= 0.2-0.3~dex except for Pb, which 
had a wider range. 
Therefore we adopted a representative scatter estimate of $\pm$0.25 for all 
elements except Pb, to which we assign $\pm$0.50.

\begin{figure}
\epsscale{0.9}
\plotone{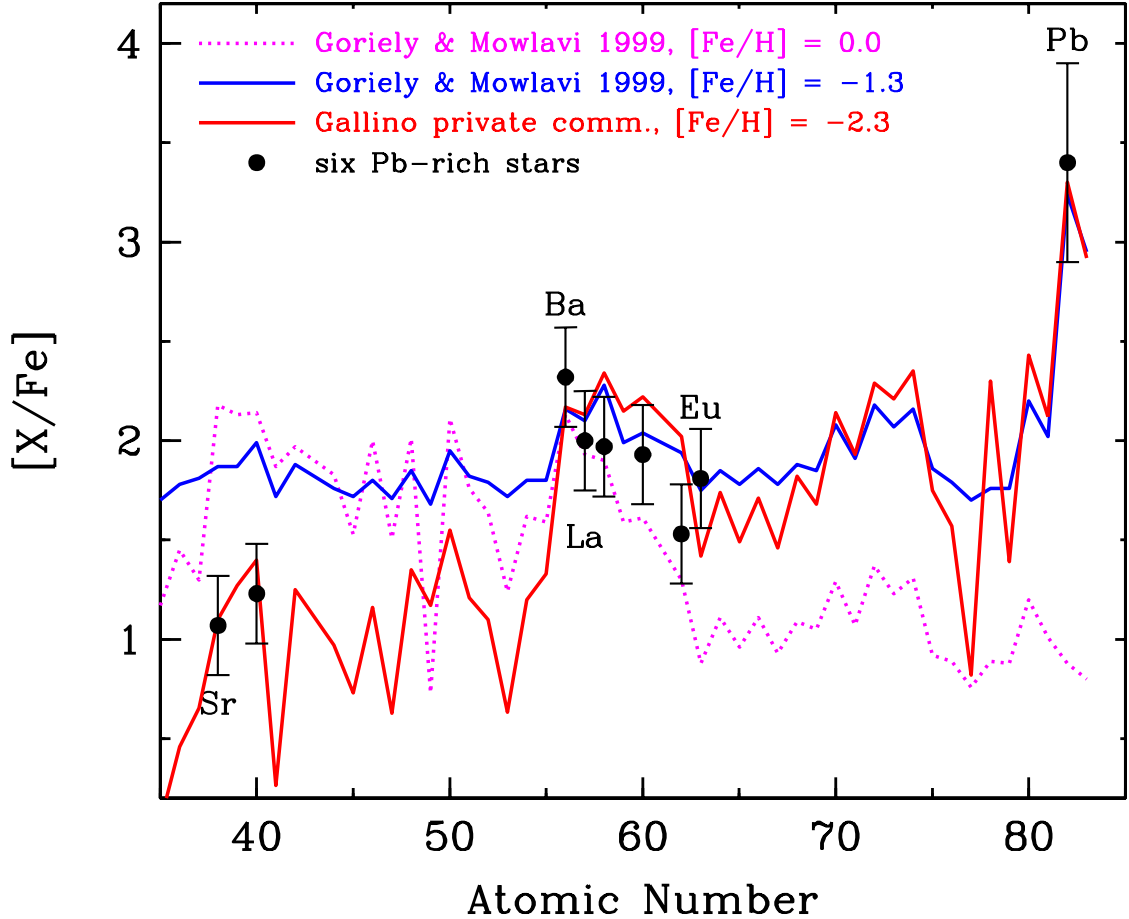}
\caption{
Comparison of the normalized mean $n$-capture abundances of the six 
Pb-rich metal-poor stars shown in the previous figure and
abundance predictions of AGB star $s$-process nucleosynthesis 
(Goriely \& Mowlavi 2000; Gallino, private communication).
Lines and symbols are defined in the figure legend.
\label{f8}}
\end{figure}

In Figure~\ref{f8} the observed mean abundances of these six Pb-rich stars
are compared with theoretical $s$-process model calculations of
Goriely \& Mowlavi (2000)\nocite{gor00} and Gallino \etal\
(2003\nocite{gal03}; private communication).
We show Goriely \& Mowlavi's $s$-process predictions arising from an
initial solar metallicity abundance distribution in the $n$-capture
synthesis region, and the predictions from their most metal-poor
([Fe/H]~= --1.3) calculation.
The Gallino \etal\ prediction is for an $s$-process synthesis environment 
with initial metallicity close to that of our BMP sample.
The major neutron-source in both sets of computations is assumed to be 
the $^{13}$C($\alpha$,n)$^{16}$O reaction.
Vertical offsets have been applied to these three theoretical curves so that 
they approximately match the mean observed Ba and La abundances.
The resulting agreement between the heavier (Z~$\geq$~56) $n$-capture 
abundances and the two metal-poor $s$-process predictions are very good.
The Gallino \etal\ distribution provides a better fit to the observed 
Sr and Zr abundances, but this may simply reflect the one dex lower
metallicity of that calculation compared to that of Goriely \& Mowlavi.

An important confirmation of these $s$-process nucleosynthesis arguments 
would be the detection of Bi in BMP binaries. 
This is the heaviest (Z~=~83) $s$-process element, and should also
be greatly enhanced in stars such as CS~29497-030. 
The Bi abundance could be as large as [Bi/Pb]~$\sim$~--0.4 to --0.3 (Goriely
\& Mowlavi 2000; Gallino private communication), or 
[Bi/Fe]~$\gtrsim$~+3 in CS~29497-030 (see Figure~\ref{f8}).
Relatively strong \ion{Bi}{1} lines exist in the near-UV and vacuum UV
spectral regions; future observational campaigns should be able to detect
these lines.

The association of large abundances of C with $s$-process production
is easy to make from the connection between triple-$\alpha$ He-burning 
and the liberation of neutrons. 
However, O should also be created in the same interior layers of an AGB 
star, both from $^{13}$C($\alpha$,n)$^{16}$O and more directly from 
$^{12}$C($\alpha$,$\gamma$)$^{16}$O.  
Production of O is very difficult to quantify, because the reaction
rate of the latter reaction is still very uncertain (\eg,
Kunz \etal\ 2002\nocite{kun02}; Heger \etal\ 2002\nocite{heg02};
Straniero \etal\ 2003\nocite{str03}, and references therein).
Evidence from our BMP sample is suggestive: the star with the most 
extreme $s$-process and C enhancements, CS~29497-030,
has an O abundance 3-4 times larger than the other stars.
Recently, Aoki \etal\ (2002a)\nocite{aok02a} have reported that
LP~625-44, a very metal-poor subgiant with large excesses of C and
the $s$-process elements, has an O overabundance that may be as
much as a factor of 10 larger than HD~140283 (a star with similar
\teff, log~$g$, and [Fe/H]).
The discussion in that paper concentrates on O synthesis via
$^{13}$C($\alpha$,n)$^{16}$O.
A large-sample comparative O abundance study between Pb-rich stars and 
those with more modest $s$-process overabundances could help quantitatively
constrain predictions of O production in metal-poor AGB stars.

There appear to be two stellar evolutionary pathways to the $s$-process
(over)production of Pb that is found in some very metal-poor main-sequence
stars.  
CS~29497-030 is probably another example of the AGB mass-transfer paradigm 
(McClure \& Woodsworth 1990\nocite{mcc90}; McClure 1997\nocite{mcc97}) 
proposed for the BMP binaries of this paper.
The other path, exemplified by CS~22880-074 and CS~22898-027 (Aoki \etal\ 
2002b\nocite{aok02b}), is not yet understood.  
The latter stars are metal-poor main sequence stars with large C and 
$s$-process overabundances, but Preston \& Sneden (2001)\nocite{pre01} 
found that they exhibited no evidence of radial velocity variations 
from 1990 to 1999.  
Aoki \etal\ confirm the continued absence of velocity
variations for both stars through 2001.

We are uncertain how to classify HE~0024-2523 into one of these groups.
Its orbital period, P~=~3.41~d (Lucatello \etal\ 2003\nocite{luc03}) lies 
far outside the period ranges found for both giant CH star binaries, 
238~$<$~P(d)~$<$~2954 (McClure \& Woodsworth 1990\nocite{mcc90}) and 
sub-giant CH star binaries, 878~$<$~P(d)~$<$~4140 
(McClure 1997\nocite{mcc97}).  
We wonder why only one of 15 CH star binaries shrank to such small orbital 
dimensions, and can imagine two alternative explanations.  
In the first, the HE~0024-2523 binary is accompanied by a third star which 
transferred mass to both of the close binary components during its AGB 
evolution.  
The separation of this putative third star from the close pair could easily 
lie within the range of known CH binary system dimensions, while exceeding
the minimum separation required for secular stability of a triple system,
whether determined empirically (Heintz 1978\nocite{hei78}) or by numerical 
simulation (Harrington 1977\nocite{har77}).  
We experimented with the published velocity data of Lucatello \etal\ and 
verified that this possibility cannot be excluded by use of the extant 
velocity data.  
We arbitrarily increased the year-2000 radial velocities by 8 \kmsec, a
value which exceeds the largest $K_1$ value for McClure's subgiant CH stars,
and find that we can successfully combine the altered data into a good 
velocity curve by making minimal adjustments to the Lucatello \etal\ orbital 
elements $P$, $K_1$, and $V_0$.\footnote{
Note that Lucatello \etal\ give 13.5 \kmsec\ as the projected synchronous 
equatorial rotational velocity for P$_{\rm orb}$~=~3.41~d and 
R~=~1.1~R$_\sun$, whereas the correct value is 16.3 \kmsec.  
This error is probably a typo, as it does not propagate into their 
calculated sin$i$.}
We conclude that careful velocity measurements for several years will be 
required to confirm or reject the triple-star hypothesis.
Batten (1973)\nocite{bat73} finds that approximately one-third of all 
binaries reside in triple systems, so this possibility cannot be dismissed 
as improbable.  
Finally, as a second possibility we state the obvious:  HE~0024-2523 could 
belong to the class of $single$ main-sequence stars with C and $s$-process 
enrichments discussed in Preston \& Sneden (2001)\nocite{pre01}. 
HE~0024-2523 might just happen to have a binary companion that has nothing 
to do with its abundance peculiarities.

\begin{figure}
\epsscale{0.9}
\plotone{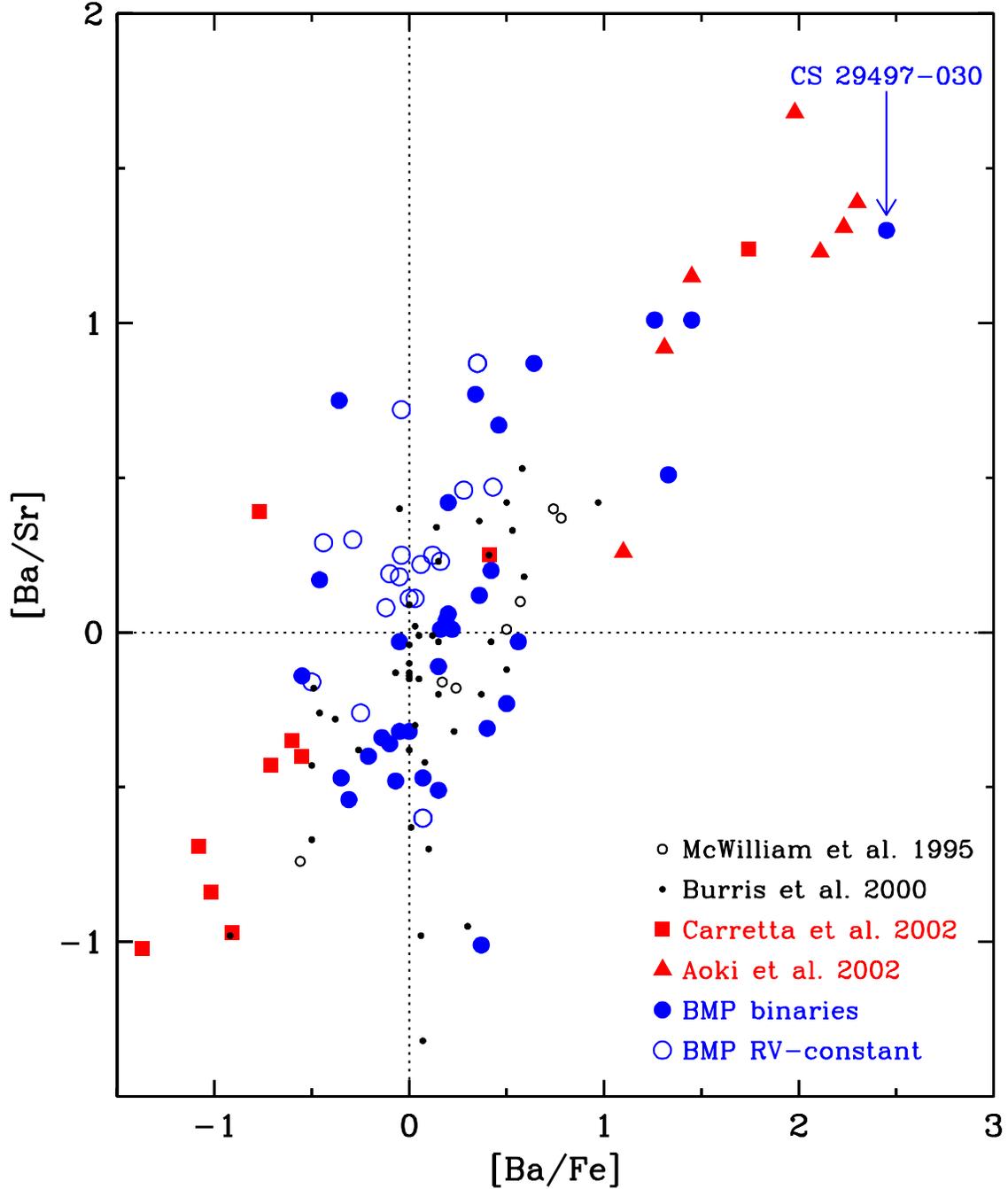}
\caption{
Correlation of [Ba/Sr] with [Ba/Fe] abundance ratios in metal-poor stars,
combining results of several surveys.  The horizontal and vertical dotted
lines indicate solar ratios of [Ba/Sr] and [Ba/Fe], respectively.
The abundances for BMP RV-constant stars (large open circles) and binaries 
(large filled circles) are taken from this study and PS00.
The McWilliam \etal\ (1995) and Burris \etal\ 
(2000) data are included because their large sample sizes
are useful in showing the domain of this diagram covered by 
ordinary metal-poor stars.
The Carretta \etal\ (2002) study includes one Pb-rich star. 
The Aoki \etal\ (2002) sample concentrated exclusively on $s$-process-rich 
stars.  
We exclude their star CS~22942-019, with its uncertain Sr abundance
based on extremely strong \ion{Sr}{2} lines.
\label{f9}}
\end{figure}

We show in Figure~\ref{f9} a comparison of the [Ba/Sr] and
[Ba/Fe] ratios for a selection of metal-poor stars. 
These include the data sets of Aoki \etal\ (2002), Carretta \etal\ (2002),
Lucatello \etal\ (2003), along with our BMP binary and RV-constant stars. 
We have also plotted data from surveys of very metal-poor Galactic 
halo stars by McWilliam \etal\ (1995) and Burris \etal\ (2000). 
Among several interesting trends illustrated in this figure, we note that 
most of the stars cluster around solar values for these ratios. 
The Burris \etal\ and McWilliam \etal\ stars shown in the figure are 
dominated by the products of the rapid $n$-capture nucleosynthesis 
(the $r$-process), rather than the $s$-process.
This is sensible since most of these metal-poor halo stars are quite old 
and formed prior to the main onset of Galactic $s$-process 
nucleosynthesis from low-mass, long stellar-evolutionary-lifetime stars.
In contrast to solar material, Ba in these stars was synthesized almost 
entirely by the $r$-process, and thus we see relatively modest ratios of
[Ba/Fe] in the surveys of McWilliam \etal\ and Burris \etal\

In the upper right quadrant of Figure~\ref{f9}, several stars have 
exceedingly high [Ba/Fe] abundance ratios -- in excess of 100 times solar. 
These stars show the result of extensive $s$-process enhancements of Ba.
Nearly all stars with [Ba/Fe]~$>$~+1 have [Ba/Sr]~$\gtrsim$~--1.
We interpret these high [Ba/Sr] ratios as a large over-production of Ba, 
rather than an under-production of Sr (see Table~7).
Even more intriguing, the stars that exhibit these large [Ba/Fe] ratios
include the most Pb-rich star (CS~29497-030) in our sample,
the one Pb-rich metal-poor star (HE~0024-2523) from Carretta \etal\ and
the most Pb-rich stars from the Aoki \etal\ study.
The one-to-one correspondence between enhanced [Ba/Fe] ratios with
very large Pb abundances offers convincing evidence for
an abundance distribution dominated by very large enhancements of the
most massive, Pb-peak, $s$-process nucleosynthesis elements. 
The extreme examples of $r$-process enhancement among 
very metal-poor stars (\eg, CS~22892-052, Sneden \etal\ 2003\nocite{sne03};
CS~31082-001, Hill \etal\ 2002\nocite{hil02}; BD~+17\deg~3248, Cowan
\etal\ 2002\nocite{cow02}), have $<$[Ba/Fe]$>$~$\approx$~+0.9 but
$<$[Ba/Sr]$>$~$\approx$~+0.3.
Therefore large [Ba/Fe] ratios accompanied by very high [Ba/Sr] ratios,
which can be easily estimated from the strong \ion{Ba}{2} and \ion{Sr}{2}
resonance lines detectable in moderate-resolution spectra,
offer simple observational search signatures for identifying 
additional Pb-rich stars.

The special case of CS~22956-028 listed in Table~5 and shown in
Figure~\ref{f9} deserves comment here.
This star has a very large [Sr/Fe] ratio but a low [Ba/Fe]
value, leading to [Ba/Sr]~$\approx$~--1.0.
We were unable to detect Pb in this star.
The large [Ba/Sr] ratio in CS~22956-028 could, therefore, be interpreted
as indicating that $s$-processing in the companion star of CS~22956-028
was much less extensive than in those cases with high Ba, and
correspondingly low [Ba/Sr] values.
Alternatively, Sr production might have been enhanced relative to Ba,
by either the weak $s$-process (\eg, Truran \etal\ 2002\nocite{tru02}) or
perhaps even some type of primary production mechanism
(\eg, Travaglio \etal\ 2003\nocite{tra03}) in massive stars.
[Ba/Sr] values $\lesssim$--1.0 are typical of $r$-process-poor stars
such as HD~122563 (Truran \etal), making it unlikely that the $r$-process
could be responsible for the synthesis of Sr in CS~22956-028.

\subsection{Intermediate-Age Stars versus Blue Stragglers among 
the Lowest Metallicity BMP stars}

We place the results presented here in the context of previous attempts to 
resolve BMP stars into intermediate-age and field blue-straggler subgroups 
(Preston \etal\ 1994; PS00). 
We restrict our attention to the 17 stars with [Fe/H]~$<$~--2 
according to PS00, summarizing their properties relevant to the present 
discussion in Table~7.
Eight of these stars possess normal or near-normal [Sr/Fe] and [Ba/Fe] 
values, and the mean values have small dispersion, as in the top portion 
of this table.
These stars may actually be Sr-deficient in the mean.  
Tabulated upper abundance limits were not used in the calculation of means.  
Two of these eight stars are single-lined spectroscopic binaries, so the 
binary fraction for this group is 0.25,
consistent with results of previous investigations of disk (Duquennoy \&
Mayor 1991\nocite{duq91}) and halo (Latham \etal\ 1998\nocite{lat98}) 
main-sequence binary frequency.  
If we include the very metal-poor double-lined spectroscopic binary 
CS~22873-139 (Preston 1994\nocite{pre94a}; Thorburn 1994\nocite{tho94}),
the binary fraction rises to 0.33, somewhat large but uncertain because of
the small database.  
We suggest that these nine stars are intermediate-age metal-poor main
sequence stars. 

The remaining eight stars, listed in the bottom portion of 
Table~7 are all spectroscopic binaries.  
The [Sr/Fe] and [Ba/Fe] values for five of them are significantly higher 
than solar, and their means exceed those of the upper portion of 
Table~7 by an order of magnitude or more, thus conforming 
to our view that the abundance peculiarities of this group arise from 
mass transfer during AGB evolution of their companions.  
Three additional binaries, also listed in the bottom portion of 
Table~7 exhibit large rotational line-broadening ($v$sin$i$~$\geq$ 35~\kmsec) 
that precluded nearly all measurements of Sr and Ba lines in our
rather noisy spectra, so we can say nothing about the element-to-Fe ratios. 
In view of their relatively short orbital periods we regard it likely that 
they experienced RGB rather than AGB mass transfer, if they are in fact 
post-mass transfer binaries.  
The very low [Sr/Fe] value of CS~29527-045 supports, but cannot prove,
this suspicion.
We identify these eight binary BMP stars as bona fide field blue 
straggler (FBS) analogs.  
If these assignments are correct the FBS fraction of the most metal-poor 
BMP stars is 8/17~=~0.47, a modest 20 percent smaller than the value 0.6 
estimated by PS00.  
Thus, the most metal-poor BMP stars are not unusual with regard to their 
makeup. 
Note that were we to assign the three rapid rotators
to the intermediate age group, the FBS fraction would fall to 5/17 = 0.29
and the intermediate-age component would become uncomfortably large 
(\eg, see Unavane, Wyse, \& Gilmore 1996\nocite{una96}).

It is surprising that all five of the FBS's with [Fe/H]$<$~--2 for
which we can measure both Sr and Ba abundances exhibit $s$-process enrichment.  
The orbital periods of these five stars, which lie in the range 
196~$\leq$~P(days)~$\leq$~1630, and the $a$sin$i$'s derived from
them in the last column of Table~7 appear to be drawn at 
random from those of the 42 binaries studied by PS00.  
None of the more metal-rich BMP binaries show significant $s$-process 
enrichment, but, as noted by our referee, a steep dependence of C and
s-process enrichment on [Fe/H] may simply be an indication that the
accreted mass fractions of these elements are approximately independent of
metallicity, as illustrated by the calculations presented in Figure~11 of
Smith, Coleman \& Lambert (1993).\nocite{smi93}
Therefore, additional AGB mass transfer binaries may be present but 
undetected among the more metal-rich BMP binary sample.  
About half of the BMP binaries with [Fe/H]~$<$~--2 avoided mass
transfer during RGB evolution (see PS00, the [Ba/Fe] panel of Figure~17), 
so we can only suppose that their large initial orbital dimensions precluded
Roche-lobe overflow, and that their orbits shrank during subsequent AGB
evolution, when Roche-lobe overflow must have occurred to provide the
observed abundance anomalies and elevated main-sequence locations of these
stars.  
We argue the case for Roche-lobe overflow rather than accretion in an AGB
superwind, because Theuns \etal\ (1996)\nocite{the96} find that wind accretion 
transfers no more than 2 to 3 percent of the wind mass from the donor to the 
receiver, an amount insufficient to move even the most massive MS 
(turnoff) stars to their FBS positions above the main sequence turnoff.  
Furthermore the effect of non-conservative mass loss by the wind
is to $increase$ both the semimajor axis and the period, not to
reduce them. 
Fortunately, a substantial number of additional BMP candidates are available 
(Preston \etal\ 1994) for further investigation of this intriguing
puzzle.

Finally, the classification scheme in Table~7 can be subjected to the 
lithium test. 
Metal-poor intermediate-age stars should lie on the Spite plateau 
(Spite \& Spite 1982\nocite{spi82}, Thorburn 1994\nocite{tho94}), 
while lithium abundances in the surface layers of FBS's should be greatly
diminished due to destruction of lithium in the red giant envelopes of 
their companions, as discussed, for example, by PS00 and Ryan \etal\ 
(2001)\nocite{rya01}.

\section{SUMMARY}

Our observations have indicated that there are distinct differences
in the abundances of BMP binaries and RV-constant stars. 
The binaries are rich in C and $n$-capture elements. 
We also suspect that this class of objects is rich in O, and if so, might
provide interesting constraints on the poorly known, but very critical,
$^{12}$C($\alpha$,$\gamma$)$^{16}$O reaction.
We have argued that most likely mass transfer from the former AGB companion
star produced the $n$-capture element abundances now observed
in the BMP binary stars.
We have found further that the $n$-capture element abundances show 
the clear signature of the $s$-process, and in particular show enhancements 
of the heaviest, Pb-peak, such elements.
Our observations have demonstrated that very abundant values of [Ba/Fe],
with correspondingly low values of [Sr/Ba], are associated directly with 
Pb-rich stars.
One of these stars, CS~29497-030 has the largest Pb abundance of any star 
yet observed. 
This would further imply that Bi abundances should be high in this
star, but we have not yet been able to confirm that.
Finally, we have drawn attention again to the class distinction between
the BMP binaries, which we identify as true field blue stragglers, and
BMP RV-constant stars, which we again suggest are intermediate-age stars
probably accreted by our Galaxy from nearby satellite dwarf spheroidals.

BMP binary stars are not alone in exhibiting C and $s$-process enhancements, 
nor are they unique in the suggested links between binary membership, 
AGB-created abundance anomalies, and mass transfer.
However, these objects represent the first stellar class for which
this particular evolutionary scenario may be ``complete''.
BMP binaries are misfits in the metal-poor HR diagram which, together with 
the abundance information presented here, clearly indicates their acquisition 
in the past of substantial amounts of envelope material from their 
companion stars.
Obviously further studies of these very interesting objects are warranted and 
will be needed to help fill in remaining parts of the story.

\acknowledgments
We thank David Lambert, Peter H\"oflich, Craig Wheeler, and the
referee for helpful discussions and suggestions for improvements to 
this paper..
Portions of this study were completed while CS was a Visiting Scientist
at the Carnegie Observatories; their hospitality and financial support
are gratefully acknowledged.
This research has been supported in part by NSF grants AST-9987162 to CS
and AST-9986974 to JJC.

\newpage

\newpage

\newpage

\newpage

\newpage

\newpage

\newpage

\newpage

\newpage

\nocite{bur00,mcw95}

\clearpage

\tablenum{1}
\tablecolumns{9}
\tablewidth{0pt}

\begin{deluxetable}{ccccccccc}

\tabletypesize{\normalsize}
\tablecaption{Basic Data and Model Atmosphere Parameters}
\tablehead{
\colhead{Star}                       &
\colhead{V}                          &
\colhead{$(B-V)_o$}                  & 
\colhead{$(U-B)_o$}                  & 
\colhead{$V_e$ sin $i$}              &
\colhead{$T_{\rm eff}$}              &
\colhead{log $g$}                    &
\colhead{$v_t$}                      &
\colhead{[Fe/H]}                     
}
\startdata
\multicolumn{4}{l}{RV-constant stars:} \\
22876-042 & 13.1 & 0.35 & --0.21 &  8 & 6750 & 4.2 & 2.50 & --2.06 \\
22941-012 & 12.5 & 0.28 & --0.17 & 10 & 7200 & 4.2 & 2.50 & --2.03 \\
22964-214 & 13.7 & 0.35 & --0.19 & 12 & 6800 & 4.5 & 2.00 & --2.30 \\
\\
\multicolumn{4}{l}{binaries:} \\
22956-028 & 13.0 & 0.34 & --0.17 & 15 & 6900 & 3.9 & 2.00 & --2.08 \\
29497-030 & 12.7 & 0.30 & --0.14 & 12 & 7050 & 4.2 & 1.75 & --2.16 \\
29509-027 & 12.5 & 0.29 & --0.13 & 10 & 7050 & 4.2 & 2.00 & --2.02 \\
\enddata

\end{deluxetable}

\clearpage

\tablenum{2}
\tablecolumns{3}
\tablewidth{0pt}

\begin{deluxetable}{@{\extracolsep{0.25in}}lrr}

\tabletypesize{\normalsize}
\tablecaption{New Radial Velocities}
\tablehead{
\colhead{Star}                       &
\colhead{JD}                         &
\colhead{RV}                         \\
\colhead{}                           &
\colhead{+2,450,000}                 &
\colhead{km s$^{-1}$}                
}
\startdata
\multicolumn{3}{l}{RV-constant stars:} \\ 
22876-042 &  2123.83  &   +42.1 \\
22876-042 &  2124.84  &   +42.4 \\
22941-012 &  2122.87  & --121.6 \\
22941-012 &  2123.80  & --122.4 \\
22941-012 &  2124.81  & --122.5 \\
22941-012 &  2474.83  & --122.5 \\
22964-214 &  2123.57  &   +41.3 \\
22964-214 &  2124.58  &   +40.8 \\
22964-214 &  2125.59  &   +41.5 \\
\\
\multicolumn{3}{l}{binaries:} \\
22956-028 &  2123.68  &   +33.6 \\
22956-028 &  2124.76  &   +33.5 \\
22956-028 &  2474.68  &   +26.5 \\
29497-030 &  2123.90  &   +48.3 \\
29497-030 &  2124.90  &   +49.1 \\
29509-027 &  2123.93  &   +77.4 \\
29509-027 &  2124.93  &   +77.4 \\
29509-027 &  2124.93  &   +77.4 \\
29509-027 &  2474.87  &   +78.1 \\
\enddata

\end{deluxetable}

\clearpage

\tablenum{3}
\tablecolumns{9}
\tablewidth{0pt}

\begin{deluxetable}{@{\extracolsep{0.10in}}ccccccccc}

\tabletypesize{\normalsize}
\tablecaption{Orbital Parameters for the Binaries}
\tablehead{
\colhead{Star}                       &
\colhead{JD$_0$}                     &
\colhead{$V_0$}                      & 
\colhead{$K_1$}                      & 
\colhead{$e$}                        &
\colhead{$\omega$}                   &
\colhead{$P$}                        &
\colhead{$\sigma_{\rm orb}$}         &
\colhead{$n$}                        \\
\colhead{}                           &
\colhead{}                           &
\colhead{km s$^{\rm -1}$}            &
\colhead{km s$^{\rm -1}$}            &
\colhead{}                           &
\colhead{}                           &
\colhead{days}                       &
\colhead{km s$^{\rm -1}$}            &
\colhead{}                           
}
\startdata
22956-028 & 48831.0 &  +34.0 &  8.5 & 0.22 & 266 & 1290 & 0.97 & 24 \\
29497-030 & 48500.0 &  +45.0 &  4.1 & 0.00 & 120 &  342 & 0.56 & 17 \\
29509-027 & 48624.0 &  +74.2 &  3.8 & 0.15 &  20 &  194 & 0.95 & 29 \\
\enddata

\end{deluxetable}

\clearpage

\tablenum{4}
\tablecolumns{5}
\tablewidth{0pt}

\begin{deluxetable}{@{\extracolsep{0.10in}}lccrc}

\tabletypesize{\normalsize}
\tablecaption{Line Data for the Abundance Analysis}
\tablehead{
\colhead{Species}                    &
\colhead{$\lambda$}                  &
\colhead{E.P.}                       &
\colhead{log $gf$}                   &
\colhead{ref.}                       \\
\colhead{}                           &
\colhead{\AA}                        &
\colhead{eV}                         &
\colhead{}                           &
\colhead{}                            
}
\startdata
\ion{C}{1}  & 5039.06 &    7.94  & --1.79   &  1 \\
\ion{C}{1}  & 5052.17 &    7.68  & --1.30   &  1 \\
\ion{C}{1}  & 5380.34 &    7.68  & --1.62   &  1 \\
\ion{C}{1}  & 5668.94 &    8.53  & --1.47   &  1 \\
\ion{C}{1}  & 6587.62 &    8.53  & --1.00   &  1 \\
\ion{C}{1}  & 7111.47 &    8.63  & --1.09   &  1 \\
\ion{C}{1}  & 7113.18 &    8.64  & --0.77   &  1 \\
\ion{C}{1}  & 7115.18 &    8.63  & --0.93   &  1 \\
\ion{C}{1}  & 7116.99 &    8.64  & --0.91   &  1 \\
\ion{O}{1}  & 7771.94 &    9.14  &  +0.37   &  1 \\
\ion{O}{1}  & 7774.17 &    9.14  &  +0.22   &  1 \\
\ion{O}{1}  & 7775.39 &    9.14  &   0.00   &  1 \\
\ion{Sr}{2} & 4077.71 &    0.00  &  +0.15   &  2 \\
\ion{Sr}{2} & 4215.52 &    0.00  & --0.17   &  2 \\
\ion{Ba}{2} & 4554.03 &    0.00  &  +0.17   &  3 \\
\ion{Ba}{2} & 5853.69 &    0.60  & --1.01   &  3 \\
\ion{Ba}{2} & 6141.73 &    0.70  & --0.08   &  3 \\
\ion{Ba}{2} & 6496.91 &    0.60  & --0.38   &  3 \\
\ion{La}{2} & 4086.71 &    0.00  & --0.07   &  4 \\
\ion{La}{2} & 4123.22 &    0.32  &  +0.13   &  4 \\
\ion{Nd}{2} & 4061.09 &    0.47  &  +0.30   &  5 \\
\ion{Eu}{2} & 4129.72 &    0.00  &  +0.22   &  6 \\
\ion{Eu}{2} & 4205.04 &    0.00  &  +0.21   &  6 \\
\ion{Pb}{1} & 4057.81 &    1.32  & --0.17   &  7 \\
\enddata

\tablerefs{1. Wiese, Fuhr, \& Deters 1996\nocite{wie96};
           2. Wiese \& Martin 1980\nocite{wie80};
           3. Gallagher 1967\nocite{gal67};
           4. Lawler, Bonvallet, \& Sneden 2001\nocite{law01a};
           5. Cowan et~al. 2002\nocite{cow02};
           6. Lawler et~al. 2001\nocite{law01b};
           7. Bi\'emont et~al. 2000
}

\end{deluxetable}

\clearpage

\tablenum{5}
\tablecolumns{16}
\tablewidth{0pt}

\begin{deluxetable}{@{\extracolsep{-0.05in}}lcccccccccccccccc}

\tabletypesize{\scriptsize}
\tablecaption{Abundances}
\tablehead{
\colhead{Star}                       &
\colhead{[Fe/H]}                     &
\colhead{[C/Fe]\tablenotemark{a}}    &
\colhead{$\sigma$}                   &
\colhead{[C/Fe]\tablenotemark{b}}    &
\colhead{$\sigma$}                   &
\colhead{\#}                         &
\colhead{[O/Fe]}                     &
\colhead{$\sigma$}                   &
\colhead{\#}                         &
\colhead{[Sr/Fe]}                    &
\colhead{$\sigma$}                   &
\colhead{\#}                         &
\colhead{[Ba/Fe]}                    &
\colhead{$\sigma$}                   &
\colhead{\#}                         \\
\colhead{}                           &
\colhead{}                           &
\colhead{CH}                         &
\colhead{}                           &
\colhead{I}                          &
\colhead{}                           &
\colhead{}                           &
\colhead{I}                          &
\colhead{}                           &
\colhead{}                           &
\colhead{II}                         &
\colhead{}                           &
\colhead{}                           &
\colhead{II}                         &
\colhead{}                           &
\colhead{}                            
}
\startdata
\multicolumn{3}{l}{RV-constant stars:} \\
CS 22876-042  & --2.06 & $<$+1.0  & 0.3  & $<$+1.0 & 0.3  & 2 & +0.54 & 
                  0.33 & 2 & --0.29 & 0.10 & 2 & --0.10 & \nodata & 1 \\
CS 22941-012  & --2.03 & $<$+1.0  & 0.3  & $<$+1.0 & 0.3  & 2 & +0.40 & 
                  0.28 & 3 & --0.07 & 0.10 & 2 &  +0.16 & \nodata & 1 \\
CS 22964-214  & --2.30 & $<$+1.0  & 0.3  & $<$+1.0 & 0.3  & 2 & +0.70 & 
                  0.04 & 2 & --0.20 & 0.25 & 2 & --0.12 & \nodata & 1 \\
mean spectrum & --2.13 & $<$+0.5  & 0.2  & $<$+0.6 & 0.18 & 6 & +0.4: & 
                  0.39 & 3 & --0.26 & 0.05 & 2 & --0.11 & \nodata & 1 \\

\multicolumn{3}{l}{binaries:} \\
CS 22956-028  & --2.08 &    +1.74 & 0.20 &   +1.34 & 0.30 & 3 & +0.5: & 
                  0.12 & 2 &  +1.38 & 0.15 & 2 &  +0.37 &    0.26 & 3 \\
CS 29497-030  & --2.16 &    +2.17 & 0.10 &   +2.11 & 0.10 & 9 & +1.13 & 
                  0.09 & 3 &  +1.15 & 0.05 & 2 &  +2.45 &    0.19 & 4 \\
CS 29509-027  & --2.02 &    +1.38 & 0.20 &   +1.38 & 0.18 & 7 & +0.55 & 
                  0.17 & 3 &  +0.82 & 0.00 & 2 &  +1.33 &    0.13 & 4 \\
mean spectrum & --2.09 &    +1.76 & 0.10 &   +1.54 & 0.22 & 7 & +0.71 & 
                  0.19 & 3 &  +1.13 & 0.04 & 2 &  +1.35 &    0.22 & 4
\enddata

\tablenotetext{a}{Relative abundance with respect to 
                  log~$\epsilon_\sun$(C)~= 8.70, derived with our line list.}

\tablenotetext{b}{Relative abundance with respect to assumed 
                  log~$\epsilon_\sun$(C)~= 8.42, derived with our line list.}

\end{deluxetable}

\clearpage

\tablenum{6}
\tablecolumns{4}
\tablewidth{0pt}

\begin{deluxetable}{lccc}

\tabletypesize{\normalsize}
\tablecaption{Abundances for CS 29497-030}
\tablehead{
\colhead{Species}                    &
\colhead{[X/Fe]\tablenotemark{a}}    &
\colhead{$\sigma$}                   &
\colhead{\#}                          
}
\startdata
CH          &     +2.17 &    0.10 & \nodata \\
\ion{C}{1}  &     +2.11 &    0.10 &       9 \\
\ion{O}{1}  &     +1.13 &    0.09 &       3 \\
\ion{Mg}{1} &     +0.47 &    0.12 &       3 \\
\ion{Ca}{1} &     +0.46 &    0.23 &       3 \\
\ion{Sc}{2} &     +0.07 &    0.34 &       3 \\
\ion{Ti}{2} &     +0.46 &    0.14 &      13 \\
\ion{Cr}{1} &    --0.26 & \nodata &       1 \\
\ion{Mn}{1} & $<$--0.14 & \nodata &       3 \\
\ion{Sr}{2} &     +1.15 &    0.05 &       2 \\
\ion{Ba}{2} &     +2.45 &    0.19 &       4 \\
\ion{La}{2} &     +1.91 &    0.21 &       2 \\
\ion{Nd}{2} &     +2.1: & \nodata &       2 \\
\ion{Eu}{2} &     +1.7: &    0.1: &       2 \\
\ion{Pb}{1} &     +3.75 & \nodata &       1 \\
\enddata

\tablenotetext{a}{Abundances of C, O, Sr, Ba, La, Nd, Eu, and Pb
                  are from this work; the other abundances are from
                  Paper~1.}

\end{deluxetable}

\clearpage

\tablenum{7}
\tablecolumns{8}
\tablewidth{0pt}

\begin{deluxetable}{lcrrrrcc}

\tabletypesize{\normalsize}
\tablecaption{Summary Properties of Lowest Metallicity BMP Stars}
\tablehead{
\colhead{Star}                       &
\colhead{Class\tablenotemark{a}}     &
\colhead{$v$sin$i$\tablenotemark{b}} &
\colhead{[Fe/H]\tablenotemark{c}}    & 
\colhead{[Sr/Fe]\tablenotemark{c}}   & 
\colhead{[Ba/Fe]\tablenotemark{c}}   &
\colhead{$P$\tablenotemark{d}}       &
\colhead{$a$sin$i$/R$_\sun$}         \\
\colhead{}                           &
\colhead{}                           &
\colhead{km s$^{-1}$}                &
\colhead{}                           &
\colhead{}                           &
\colhead{}                           &
\colhead{days}                       &
\colhead{}                           
}
\startdata
\multicolumn{5}{l}{Intermediate Age:} \\
CS 22873-139 & SB  & 10 & --2.85 & \nodata &   \nodata &       19 &       9 \\
CS 22876-042 & RVC &  8 & --2.06 &  --0.29 &    --0.10 &  \nodata & \nodata \\
CS 22880-013 & RVC & 12 & --2.05 &  --0.59 & $<$--0.29 &  \nodata & \nodata \\
CS 22941-005 & SB  & 25 & --2.43 &  --0.43 &  $<$+0.34 & 324,3000 &  42,382 \\
CS 22941-012 & RVC & 10 & --2.03 &  --0.07 &     +0.16 &  \nodata & \nodata \\
CS 22950-173 & RVC &  8 & --2.50 &  --0.76 &    --0.04 &  \nodata & \nodata \\
CS 22960-058 & RVC & 10 & --2.13 &  --0.13 &     +0.12 &  \nodata & \nodata \\
CS 22964-214 & RVC & 12 & --2.30 &  --0.20 &    --0.12 &  \nodata & \nodata \\
CS 29499-057 & SB  & 25 & --2.33 &  --0.23 &  $<$+0.64 &     2500 &      92 \\
mean         &     &    &        &  --0.34 &      0.00 &          &         \\
$\sigma$     &     &    &        &    0.24 &      0.13 &          &         \\
\\
\multicolumn{5}{l}{Field Blue Straggler:} \\
CS 22890-069 & SB  & 80 & --2.00 & \nodata &   \nodata &        2 &     1.3 \\
CS 22946-011 & SB  & 15 & --2.59 &   +0.25 &     +1.26 &      585 &     126 \\
CS 22956-028 & SB  & 15 & --2.08 &   +1.38 &     +0.37 &     1290 &     210 \\
CS 22963-013 & SB  & 70 & --2.50 & \nodata &   \nodata &       85 &      21 \\
CS 29497-030 & SB  & 12 & --2.16 &   +1.15 &     +2.45 &      342 &      25 \\
CS 29509-027 & SB  & 10 & --2.01 &   +0.82 &     +1.33 &      194 &      14 \\
CS 29518-039 & SB  & 17 & --2.49 &   +0.44 &     +1.45 &     1630 &     254 \\
CS 29527-045 & SB  & 35 & --2.14 &  --0.34 &   \nodata &       84 &      27 \\
mean         &     &    &        &   +0.62 &     +1.37 &          &         \\
$\sigma$     &     &    &        &    0.63 &      0.74 &          &         \\
\enddata

\tablenotetext{a}{RVC = RV-constant star; SB = spectroscopic binary.}

\tablenotetext{b}{PS00, Table~1.}

\tablenotetext{c}{This paper, Table~5, otherwise PS00, Table~7;
                  for CS~22890-069 and CS~22963-013 [Fe/H] estimates
                  from K-line strength are from PS00, Table~1.}

\tablenotetext{d}{This paper, Table~3, otherwise PS00, Table~5.}

\end{deluxetable}


\clearpage 

\begin{thebibliography}

\bibitem[]{aok01}
Aoki, W., Ryan, S. G., Norris, J. E., Beers, T. C., Ando, H., Iwamoto, N., 
Kajino, T., Mathews, G. J., \& Fujimoto, M. Y. 2001, \apj, 561, 346

\bibitem[]{aok02a}
Aoki, W. et al. 2002a, \pasj, 54, 427

\bibitem[]{aok02b}
Aoki, W., Ryan, S. G., Norris, J. E., Beers, T. C., Ando, H., 
\& Tsangarides, S., 2002b, \apj, 580, 1149

\bibitem[]{bat73}
Batten, A. 1973, Binary and Multiple Systems of Stars (Pergamon Press:
Oxford), 62

\bibitem[]{bie00}
Bi\'emont, E., Garnir, H. P.; Palmeri, P., Li, Z. S., \& Svanberg, S.
2000, \mnras, 312, 116

\bibitem[]{bur00}
Burris, D. L., Pilachowski, C. A., Armandroff, T. A., Sneden, C.,
Cowan, J. J., \& Roe, H. 2000, \apj, 544, 302

\bibitem[]{bus99}
Busso, M., Gallino, R., \& Wasserburg, G. J. 1999, \apj, \araa, 37, 239

\bibitem[]{car02}
Carretta, E., Gratton, R., Cohen, J. G., Beers, T. C., \& Christlieb, N.
2002, \aj, 124, 481

\bibitem[]{cow02}
Cowan, J. J., et al. 2002, \apj, 572, 861

\bibitem[]{duq91}
Duquennoy, A., \& Mayor, M. 1991, \aap, 248, 485

\bibitem[]{gal67}
Gallagher, A. 1967, Phys. Rev., 157, 24

\bibitem[]{gal03}
Gallino, R., et al. 2003, in Proc. 7th Int. Conf. on Nuclei in the Cosmos,
in press

\bibitem[]{gor00}
Goriely, S., \& Mowlavi, N. 2000, \aap, 362, 599

\bibitem[]{har77}
Harrington, R. S. 1977, Rev. Mex. Ast. \& Ap., 3, 139

\bibitem[]{heg02}
Heger, A., Woosley, S. E., Rauscher, T., Hoffman, R. D., \& Boyes, M. M.
2002, New Ast. Rev., 46, 463

\bibitem[]{hei78}
Heintz, W. D. 1978, Double Stars, Geophysics \& Astrophysics Monographs,
Vol. 15 (Reidel: Dordrecht), 66

\bibitem[]{hil02}
Hill, V. \etal\ 2002, \aap, 387, 560

\bibitem[]{hol74}
Holweger, H. \& M{\"u}ller, E. A.  1974, \solphys, 39, 19

\bibitem[]{joh02}
Johnson, J. A., \& Bolte, M. 2002, \apj, 579, L87

\bibitem[]{kun02}
Kunz, R., Fey, M., Jaeger, M., Mayer, A., Hammer, J. W., Staudt, G.,
 Harissopulos, S., \& Paradellis, T. 2002, \apj, 567, 643

\bibitem[]{kur95}
Kurucz, R. L. 1995, in Workshop on Laboratory and astronomical high
resolution spectra, ASP Conference Ser. \#81 ed. A.J. Sauval, R. Blomme,
and N. Grevesse (San Francisco: Astr. Soc. Pac.), p.583

\bibitem[]{kur84}
Kurucz, R. L., Furenlid, I., Brault, J., \& Testerman, L.  1984,
Solar Flux Atlas from 296 to 1300 nm (Cambridge, MA: Harvard Univ.)

\bibitem[]{lat98}
Latham, D. W., Stefanik, R. P., Mazeh, T., Goldberg, D., Torres, G., 
\& Carney, B. W. 1998, in Cool Stars, Stellar Systems, and the Sun, ASP Conf. 
Ser. 154, ed. R. A. Donahue \& J. A. Bookbinder (San Fransisco: ASP), 2129

\bibitem[]{law01a}
Lawler, J. E., Bonvallet, G., \& Sneden, C. 2001, \apj, 556, 452

\bibitem[]{law01b}
Lawler, J. E., Wickliffe, M. E., den Hartog, E. A., \& Sneden, C.
2001, \apj, 563, 1075

\bibitem[]{luc03}
Lucatello, S., Gratton, R., Cohen, J. G., Beers, T. C., Christlieb, N.,
Carretta, E., \& Ramirez, S. 2003, \aj, 125, 875

\bibitem[]{mcc90}
McClure, R. D. 1997, \pasp, 109, 536

\bibitem[]{mcc97}
McClure, R. D., \& Woodsworth, A. W. 1990, \apj, 352, 709

\bibitem[]{mcc64}
McCrea, W. H. 1964, \mnras, 128, 147

\bibitem[]{mcw95}
McWilliam, A., Preston, G. W., Sneden, C., \& Searle, L.  1995,
\aj, 109, 2757

\bibitem[]{pre94a}
Preston, G. W. 1994, \aj, 108, 2267

\bibitem[]{pre94}
Preston, G. W., Beers, T. C., \& Shectman, S. A. 1994, \aj, 108, 538

\bibitem[]{pre00}
Preston, G. W., \& Sneden, C. 2000, \aj, 120, 1014 (PS00)

\bibitem[]{pre01}
Preston, G. W., \& Sneden, C. 2001, \aj, 122, 1545

\bibitem[]{ros99}
Rossi, S., Beers, T. C., \& Sneden, C. 1999, in The Third Stromlo Symposium: 
the Galactic Halo, ed. B. K. Gibson, T. S. Axelrod, M. E. Putman, ASP
Conf. Ser., 165, 264

\bibitem[]{rya01}
Ryan, S. G., Beers, T. C., Kajino, T., Rosolankova, K. 2001, \apj, 547, 231

\bibitem[]{she84}
Shectman, S. A. 1984, in Proc. SPIE 445, Instrumentation in Astronomy V,
ed. A. Boksenberg \& D. L. Crawford (Bellingham: SPIE), 128

\bibitem[]{sil99}
Sills, A., \& Bailyn, C. D. 1999, \apj, 513, 428

\bibitem[]{sme94}
Smecker-Hane, T., Stetson, P., B., Hesser, J. E., \& Lennert, M. D. 
1994, \aj, 108, 507

\bibitem[]{smi93}
Smith, V. V., Coleman, H., Lambert, \& D. L. 1993, \apj, 417, 287

\bibitem[]{sne73}
Sneden, C. 1973,  \apj, 184, 839

\bibitem[]{sne03}
Sneden, C. 2003, \apj, submitted

\bibitem[]{spi82}
Spite, F., \& Spite, M. 1982, \aap, 115, 357

\bibitem[]{str03}
Straniero, O., Domnguez, I., Imbriani, G., \& Piersanti, L. 2003,
\apj, 583, 878

\bibitem[]{str93}
Stryker, L. L. 1993, \pasp, 105, 1081

\bibitem[]{the96}
Theuns, T., Boffin, H. M., \& Jorrisen, A. 1996, \mnras, 2890, 1264

\bibitem[]{tho94}
Thorburn, J. A. 1994, \apj, 421, 318

\bibitem[]{tra03}
Travaglio, C., Gallino, R., Cowan, J. J., Jordan, F., \& Sneden, C.
2003, \apj, to be submitted

\bibitem[]{tru02}
Truran, J. W.,  Cowan, J. J., Pilachowski, C. A., \& Sneden, C. 2002,  
\pasp, 114, 1293

\bibitem[]{una96}
Unavane, M., Wyse, R. F. G., \& Gilmore, G. 1996, \mnras, 278, 727

\bibitem[]{van01}
Van Eck, S., Goriely, S., Jorissen, A., \& Plez, B. 2001, \nat, 412, 793

\bibitem[]{van03}
Van Eck, S., Goriely, S., Jorissen, A., \& Plez, B. 2003, preprint

\bibitem[]{vil82}
Vilhu, O. 1982, \aap, 109, 17

\bibitem[]{wie96}
Wiese, W. L., Fuhr, J. R., \& Deters, T. M. 1996, J. Phys. Chem. Ref. 
Data Mono. 7 

\bibitem[]{wie80}
Wiese, W. L., \& Martin, G. A. 1980, Wavelengths and Transition Probabilities
for Atoms and Atomic Ions: Part II, NBS-NSRDS, No. 68

\end{thebibliography}
\end{document}